\documentclass[prl,aps,superscriptaddress,floatfix,tightenlines,showpacs,twocolumn]{revtex4-2}
\usepackage{graphicx}
\usepackage{dcolumn}   
\usepackage{bm}        
\usepackage{amssymb}   
\usepackage{amsmath}
\usepackage{url}
\usepackage{layout}
\usepackage{color}
\usepackage{epsfig}
\usepackage{graphicx}
\usepackage{mathrsfs}
\usepackage{bm}
\usepackage{appendix}
\usepackage{color}

\usepackage[thinlines]{easytable}
\usepackage{makecell}
\usepackage{tikz,pgf}
\usepackage{booktabs}
\usepackage{float}
\usepackage{hyperref}
\hypersetup{colorlinks,allcolors=blue}

\def\be{\begin{equation}}       \def\ee{\end{equation}}
\def\bea{\begin{eqnarray}}      \def\eea{\end{eqnarray}}
\def\bp{\begin{pmatrix}} \def\ep{\end{pmatrix}}
\def\beaa{\begin{equation}\begin{aligned}}
		\def\eeaa{\end{aligned}\end{equation}}

\begin{document}

\title{Superconductivity in La$_4$Ni$_3$O$_{10}$ Under Pressure}

\author{Chen Lu}
\thanks{These two authors contributed equally to this work.}
\affiliation{School of Physics, Hangzhou Normal University, Hangzhou 311121, China}
\author{Zhiming Pan}
\thanks{These two authors contributed equally to this work.}
\affiliation{Department of Physics, Xiamen University, Xiamen 361005, China}
\author{Fan Yang}
\email{yangfan\_blg@bit.edu.cn}
\affiliation{School of Physics, Beijing Institute of Technology, Beijing 100081, China}
\author{Congjun Wu}
\email{wucongjun@westlake.edu.cn}
\affiliation{New Cornerstone Science Laboratory, Department of Physics, School of Science, Westlake University, Hangzhou 310024, Zhejiang, China}
\affiliation{Institute for Theoretical Sciences, Westlake University, Hangzhou 310024, Zhejiang, China}
\affiliation{Key Laboratory for Quantum Materials of Zhejiang Province, School of Science, Westlake University, Hangzhou 310024, Zhejiang, China}
\affiliation{Institute of Natural Sciences, Westlake Institute for Advanced Study, Hangzhou 310024, Zhejiang, China}

\begin{abstract}
The discovery of superconductivity (SC) in the trilayer nickelate compound La$_{4}$Ni$_3$O$_{10}$ under pressure has generated significant interest. 
In this work, we propose a trilayer two $E_g$-orbital $t$-$J_{\parallel}$-$J_{\perp}$ model to investigate the microscopic origin of SC in this system. 
In the strong-coupling regime, each layer is governed by a $t$-$J_{\parallel}$ model with intra-layer antiferromagnetic exchange $J_{\parallel}$, while electrons are allowed to hop between layers, interacting via inter-layer exchange $J_{\perp}$. 
The inner-layer $3d_{z^2}$-orbital electrons tends to form bonding states with those in the neighboring layers, leading to redistribution of the electron densities. 
The numerical simulation results indicate that SC is predominantly mediated by the $3d_{z^2}$ orbital, characterized by an intra-layer extended $s$-wave pairing in the outer layers, accompanied by an inter-layer pairing with opposite sign. 
Furthermore, we find that electron doping enhances SC, while hole doping tends to suppress it. 
These findings provide new insights into the SC mechanisms of La$_{4}$Ni$_3$O$_{10}$ and its sensitivity to charge doping.
\end{abstract}

\maketitle


{\bf Introduction:} The Ruddlesden-Popper phase nickelates series La$_{n+1}$Ni$_n$O$_{3n+1}$ has attracted significant attention for a long time due to its potential for hosting unconventional superconductivity (SC) \cite{taniguchi1995transport,seo1996electronic,kobayashi1996transport,greenblatt1997ruddlesden,greenblatt1997electronic,ling2000neutron,wu2001magnetic,fukamachi2001nmr,voronin2001neutron,Bannikov2006,hosoya2008pressure,pardo2011dft,nakata2017finite,mochizuki2018strain,li2020epitaxial,song2020structure,barone2021improved,Wang2022LNO}.
A major breakthrough occurred with the discovery of SC in the infinite-layer Nd$_{0.8}$Sr$_{0.2}$NiO$_{3}$ ($n=\infty$) \cite{li2019nickelate}.
More recently, high-temperature SC ($T_c\approx 80$K) was observed in La$_3$Ni$_2$O$_7$ ($n=2$) under pressure exceeding $14$GPa \cite{Wang2023LNO,Wang2023LNOb,YuanHQ2023LNO,wang2023LNOpoly,wang2023la2prnio7}, sparking further investigations.
Further excitement was generated by the successful realization of SC in thin-film La$_3$Ni$_2$O$_7$ under ambient pressure, exhibiting $T_c$ approximately $40$K \cite{ko2024signatures}.
Additionally, experimental evidence of SC ($T_c\approx 20$K) under pressure in La$_4$Ni$_3$O$_{10}$ ($n=3$) \cite{li2023trilayer,zhu2023trilayer,zhang2023trilayer} has drawn considerable attention \cite{Yuan2024la3,sakakibara2023trilayer,li2024la3,kakoi2023multiband,leonov2024la3,tian2024effective,wang2024nonfermi,labollita2024electronic,zhang2024prediction,yang2024effective,luo2024trilayer,zhang2024s,li2024ultrafast,lechermann2024electronic,huang2024interlayer,oh2024type,qin2024frustrated,xu2024origin,du2024correlated,huo2024electronic,yang2024decomposition,zhang2024magnetic,huang2024signature,liu2024evolution,deswal2024dynamics,zhao2024electronic}.

The potential SC in La$_{n+1}$Ni$_n$O$_{3n+1}$ is intricately related to the electronic properties of the NiO$_2$ planes,
contributing from the Ni $E_g$ orbitals.
In single NiO$_2$ layer, the two $E_g$ orbitals are nearly degenerate.
However, in a multilayer system, inter-layer hopping lifts this degeneracy by forming bonding and anti-bonding bands.
Based on the density functional theory (DFT) calculations \cite{pardo2011dft,Wang2023LNO,YaoDX2023}, a bilayer two-orbital model has been proposed to explain the high-$T_c$ SC in La$_3$Ni$_2$O$_7$ \cite{lu2023bilayertJ,lu2023interplay}, 
where the bonding band of the $3d_{z^2}$ orbital is nearly fully occupied and localized.
In the strong coupling limit, the $3d_{z^2}$ orbital exhibits robust inter-layer superexchange, which is transmitted to the $3d_{x^2-y^2}$-orbital under strong Hund's rule \cite{lu2023bilayertJ,oh2023type2,lu2023interplay,tian2023correlation,qu2023roles,kakoi2023pair}, 
significantly enhancing the inter-layer $s$-wave SC \cite{lu2023bilayertJ,qu2023bilayer,oh2023type2,lu2023interplay,chen2023iPEPS,tian2023correlation,qu2023roles,kakoi2023pair}.

In La$_4$Ni$_3$O$_{10}$, the trilayer structure adds further complexity to its electronic nature \cite{sakakibara2023trilayer,li2017fermiology}.
Under pressure, trilayer La$_4$Ni$_3$O$_{10}$ undergoes a structural transition to the tetragonal $I4/mmm$ phase \cite{li2024la3}.
Valence counting suggests an average electronic configuration of  Ni$^{2.67+}$ ($3d^{7.33}$), 
assigning four electrons to the two $E_g$ orbitals across the three Ni atoms in each trilayer unit.
Without considering the inter-layer hopping, inter-orbital hybridization or correlation effects, the hole number per $E_g$ orbital is approximately $0.33$, placing the system in the overdoped regime, 
according to cuprate standards \cite{kotliar1988,lee2006htsc,keimer2015highTc,proust2019highTc}.
Unlike the bilayer case, both $E_g$ orbitals act as itinerant carriers, and the three layers are not fully equivalent, 
raising the fundamental question: 
what is the nature of SC in trilayer La$_4$Ni$_3$O$_{10}$?

In this study, we develop a strongly-correlated two $E_g$-orbital $t$-$J_{\parallel}$-$J_{\perp}$ model to explore the SC in trilayer La$_4$Ni$_3$O$_{10}$ under pressure. 
Employing slave-boson mean-field (SBMF) theory \cite{kotliar1988,lee2006htsc}, we self-consistently estimate the ground-state order parameters and holon densities. 
The trilayer structure, combined with finite hybridization effect, leads to itinerant $E_g$ orbital bands with effective hole densities in the overdoped regime.
Our numerical simulations show that SC in this trilayer system is predominantly driven by the $3d_{z^2}$ orbital, manifesting as intra-layer extended $s$-wave pairing in the outer layers, accompanied by inter-layer pairing.
In contrast, the $3d_{x^2-y^2}$ orbital exhibits much weaker pairing strength.
We further explore the effect of doping to the La$_4$Ni$_3$O$_{10}$.
Namely, we find that electron doping enhances SC, while hole doping suppresses it.
Our result provide additional insights into the SC mechanism of trilayer La$_4$Ni$_3$O$_{10}$.

\begin{figure}[t!]
\centering
\includegraphics[width=0.48\textwidth]{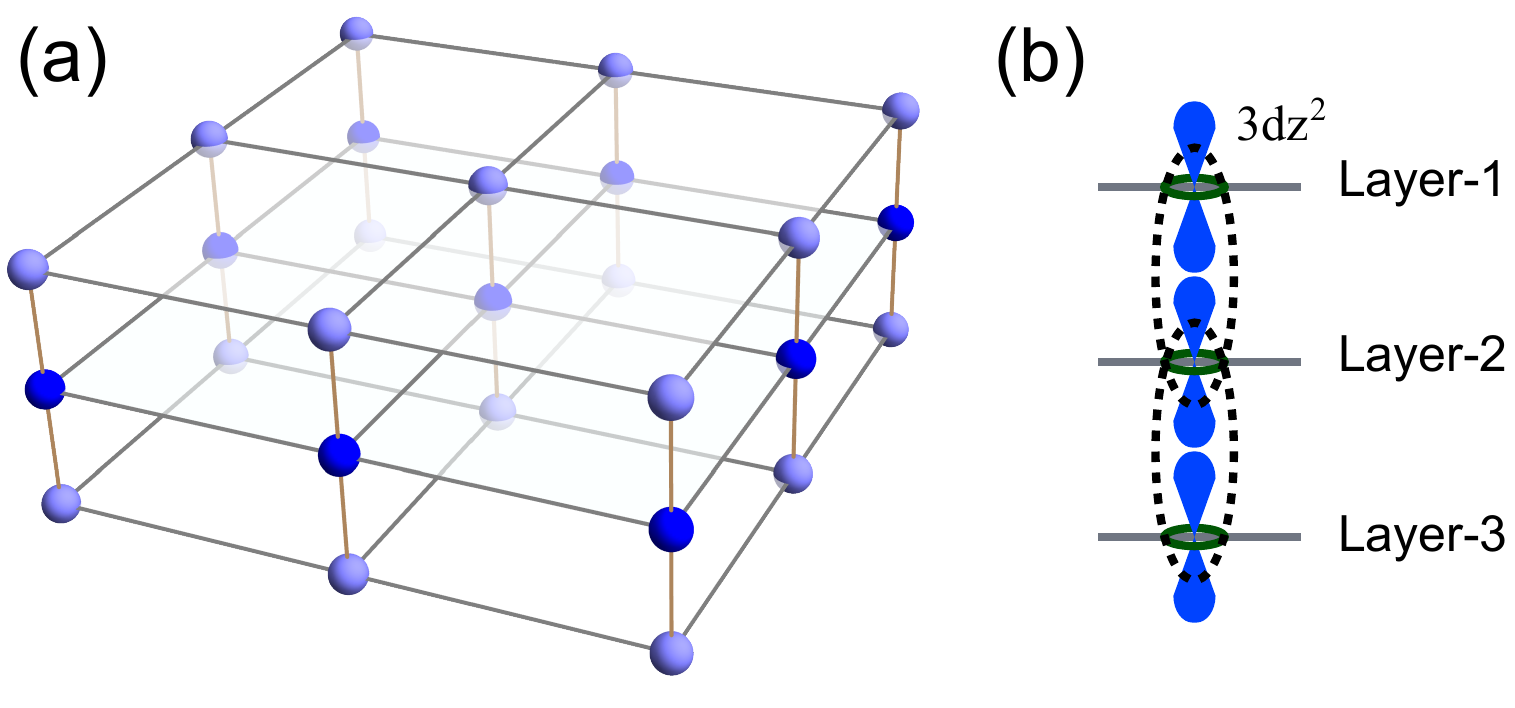}
\caption{(a) Trilayer lattice structure for the La$_3$Ni$_2$O$_7$.  
(b) Formation of the bonding bands in $3d_{z^2}$ orbital. 
The inner-layer 2 could form bonding band
with both the outer-layer 1 and 3.}
\label{fig:TrilayerLattice}
\end{figure}


{\bf Effective trilayer two-orbital model:}
The electronic characteristics of the trilayer La$_4$Ni$_3$O$_{10}$ under pressure are predominantly influenced by the two $E_g$ orbitals within the NiO$_2$ planes.
The trilayer configuration of the Ni ions is schematically shown in Fig.~\ref{fig:TrilayerLattice}(a).
Due to inter-layer hopping, $3d_{z^2}$-orbital electrons residing on the inner layer can form bonding bands with those in both the upper and lower layers, 
as depicted in Fig.~\ref{fig:TrilayerLattice}(b), which may be viewed as geometric frustration.
Consequently, $3d_{z^2}$-orbital hole numbers within inner and outer layers would diverge.
The electronic nature within inner and outer layers would be different.
Experimentally, evidence points towards correlation effects, including observations of strange metal behavior in the normal state \cite{zhu2023trilayer} and suppressed kinetic energy indicated by optical studies \cite{xu2024origin,liu2024evolution}. 
Furthermore, the local Hubbard interaction $U$ on the Ni ions, largely insensitive to the specific lattice details, is estimated to be $U\approx 4-5$eV (e.g., a similar value was used for La$_3$Ni$_2$O$_7$ in \cite{pardo2011dft}), significantly exceeding the characteristic inter-site hopping energy $0.7$eV. 
These points firmly place the system within the correlated regime, thereby motivating the development of a strong-coupling effective theory in the subsequent analysis.

The intra-layer Hamiltonian $H_{\parallel}$ is given by, 
\begin{equation}
\label{two_orbital_parallel}
\begin{aligned}
&H_{\parallel}
=-\sum_{\langle i,j \rangle,\alpha,\sigma}
\big(t_{\parallel}^{x\alpha}  d_{x\alpha\sigma}^{\dagger} (i) d_{x\alpha\sigma} (j) +\text{h.c.}  \big)  \\
&-\sum_{\langle i,j \rangle,\alpha,\sigma} \big( t_{\parallel}^{z\alpha}  d_{z\alpha\sigma}^{\dagger} (i) d_{z\alpha\sigma} (j) +\text{h.c.} \big)  \\
&-\sum_{\langle i,j\rangle ,\alpha,\sigma} t_{\parallel,j-i}^{xz\alpha} \big(d_{x\alpha\sigma}^{\dagger}(i) d_{z\alpha\sigma}(j) +\text{h.c.}\big)  \\
&+ \sum_{\langle i,j \rangle,\alpha} 
\Big[ J_{\parallel}^{x\alpha}\bm{S}_{x\alpha}(i) \cdot\bm{S}_{x\alpha}(j)   
+ J_{\parallel}^{z\alpha} \bm{S}_{z\alpha}(i) \cdot\bm{S}_{z\alpha}(j)  \Big],
\end{aligned}
\end{equation}
where $d_{x\alpha\sigma}^{\dagger}(i)/d_{z\alpha\sigma}^{\dagger}(i)$ creates a $3d_{x^2-y^2}$/$3d_{z^2}$-orbital electron with spin $\sigma=\uparrow,\downarrow$ at the lattice site $i$ in the layer $\alpha=1,2,3$. 
$\bm{S}_{x\alpha}(i)=\frac{1}{2}d_{x\alpha\sigma}^{\dagger}(i) [\bm{\sigma}]_{\sigma\sigma^{\prime}}d_{x\alpha\sigma^{\prime}}(i)$ is the spin operator for the $3d_{x^2-y^2}$ orbital, with Pauli matrix $\bm{\sigma}=(\sigma_x,\sigma_y,\sigma_z)$;
similar for the $3d_{z^2}$ orbital spin $\bm{S}_{z\alpha}(i)$.
$\langle {i}{j}\rangle$ represents the summation over all the intra-layer nearest-neighbor (NN) bonds.
$H_{\parallel}$ describes three intra-layer $t$-$J_{\parallel}$ models of the two $E_g$ orbitals, with the corresponding intra-layer NN hopping terms, $t_{\parallel}^{x\alpha}$ and $t_{\parallel}^{z\alpha}$, and antiferromagnetic (AFM) spin-exchange terms, $J_{\parallel}^{x\alpha}$ and $J_{\parallel}^{z\alpha}$. 
$t_{\parallel,j-i}^{xz\alpha}$ represent the finite NN hybridization between the two $E_g$ orbitals, 
which exhibit opposite signs along $x$ and $y$ directions due to the symmetry constraint, $t_{\parallel,x}^{xz\alpha}=-t_{\parallel,y}^{xz\alpha}$.
The hole densities within $\alpha$ layer are denoted as $\delta_{x\alpha}$ and $\delta_{z\alpha}$ for $3d_{x^2-y^2}$ and $3d_{z^2}$ orbitals, respectively.

The interactions among the three layers are facilitated by the inter-layer hoppings of $3d_{z^2}$ orbitals and inter-layer superexchanges.
The three layers are categorized into two types, with the inner layer denoted by $\alpha=2$ and the outer layers labeled as $\alpha=1,3$. 
The inter-layer hoppings and exchanges are described by the following expressions:
\begin{equation}
\label{two_orbital_perp}
\begin{aligned}
&H_{\perp}
=-\sum_{i,\sigma;\alpha=1,3}
\big( t_{\perp}^z  d_{z2\sigma}^{\dagger} (i) d_{z\alpha\sigma} (i) +\text{h.c.} \big)  \\
&+ \sum_{i;\alpha=1,3}
\Big[ J_{\perp}^x\bm{S}_{x2}(i) \cdot\bm{S}_{x\alpha}(i) 
+ J_{\perp}^z \bm{S}_{z2}(i) \cdot\bm{S}_{z\alpha}(i)  \Big],
\end{aligned}
\end{equation}
involving the NN inter-layer hoppings $t_{\perp}^z$ and AFM spin-exchanges between the inner layer $\alpha=2$ and the outer layers $\alpha=1,3$.
The robust inter-layer exchange $J_{\perp}^z$ arises from the NN inter-layer hopping of the $3d_{z^2}$ orbital in the strong coupling limit. 
Additionally, a finite superexchange $J_{\perp}^x$ associated with the $3d_{x^2-y^2}$ orbital is transmitted from the $3d_{z^2}$ orbital due to the robust Hund's rule coupling \cite{lu2023bilayertJ,oh2023type2,lu2023interplay,tian2023correlation,qu2023roles,kakoi2023pair}, as depicted in Fig.~\ref{fig:TrilayerLattice}(b). 
It should be noted that this mechanism can only be realized when the $3d_{z^2}$ orbital is singly occupied, resulting in the effective renormalized $J_{\perp}^x$ by the filling of $3d_{z^2}$ orbital.
A similar situation leads to a non-zero effective intra-layer superexchange $J_{\parallel}^{z\alpha}$ for the $3d_{z^2}$ orbital, which is transmitted from the $3d_{x^2-y^2}$ orbital.
A preliminary analysis yields the following leading-order approximations,
\begin{equation}
\begin{aligned}
J_{\parallel}^{z1}=&J_{\parallel}^{z3}
=J_{\parallel}(1-\delta_{x1})^2,  \\
J_{\parallel}^{z2}=&J_{\parallel}(1-\delta_{x2})^2,  \\
J_{\perp}^{x}=&J_{\perp} (1-\delta_{z1}) (1-\delta_{z2}),  
\end{aligned}
\label{eq:EffTransJ}
\end{equation}
where the effective exchanges are mostly generated when the intermediate orbitals are occupied.
The larger holon densities would suppress these effective AFM spin exchanges.

\begin{table}[t]
	\centering
	\begin{TAB}(r,0.05cm,0.1cm)[2pt]{|c|c|c|c|c|}{|c|c|c|c|c|c|c|c|c|c|}
		OP &  GS value (eV) & & 
		OP &  GS value (eV) \\
		$\Delta_{\perp}^{z}$  & $-2.97\times 10^{-4}$
		& & $\chi_{\perp}^{z}$ &$1.17\times 10^{-1}$ \\
		$\Delta_{\perp}^{x}$  & $-1.0\times 10^{-5}$
		& & $\chi_{\perp}^{x}$ & $1.22\times 10^{-3}$ \\
		$\Delta_{\parallel}^{x2}$  & $4.37\times 10^{-5}$
		& & $\chi_{\parallel}^{x2}$ & $2.79\times 10^{-2}$ \\
		$\Delta_{\parallel}^{z2}$  & $2.73\times 10^{-5}$
		& & $\chi_{\parallel}^{z2}$ & $4.07\times 10^{-4}$ \\
		$\Delta_{\parallel}^{x1}$  & $4.33\times 10^{-5}$
		& & $\chi_{\parallel}^{x1}$ & $3.10\times 10^{-2}$ \\
		$\Delta_{\parallel}^{z1}$  & $5.28\times 10^{-4}$
		& & $\chi_{\parallel}^{z1}$ & $1.32\times 10^{-3}$ \\
		&  Holon density
		& &  & Holon density \\
		$\delta _{x1}$ &  $0.402$
		& & $\delta _{z1}$ & $0.171$ \\
		$\delta _{x2}$ &  $0.523$
		& & $\delta _{z2}$ & $0.332$ \\
	\end{TAB}
	\caption{Table of the hopping and pairing order parameters (OPs) as well as the holon densities $\delta_{x\alpha}$ and $\delta_{z\alpha}$ calculated by the SBMF theory in the ground state (GS) for $J_{\parallel}=0.25$eV and		$J_{\perp}/J_{\parallel}=2$.}
	\label{tab:MForderParameter}
\end{table}

We adopt the physical parameters of the hopping parameters from the DFT calculations with full structure relaxations \cite{zhang2024s}. 
The hopping parameters within the outer-layers are given by $t_{\parallel}^{x1}=t_{\parallel}^{x3}=0.532$eV, $t_{\parallel}^{z1}=t_{\parallel}^{z3}=0.1558$eV and $t_{\parallel}^{xz1}=t_{\parallel}^{xz3}=0.282$eV.
For the inner-layer, $t_{\parallel}^{x2}=0.545$eV, $t_{\parallel}^{z2}=0.1373$eV and $t_{\parallel}^{xz2}=0.2945$eV.
The robust inter-layer hopping for the $3d_{z^2}$ orbital is $t_{\perp}^z=0.7082$eV, while the inter-layer one for $3d_{x^2-y^2}$ nearly vanishes. 
The on-site energies are chosen as $E_{x1}=E_{x3}=0.358$eV, $E_{x2}=0.656$eV, $E_{z1}=E_{z3}=0$eV and $E_{z2}=0.42$eV.

In the strong coupling limit, 
effective AFM spin exchanges can arise from the NN hoppings.
We take the inter-layer AFM strength for $3d_{z^2}$-orbital as $J_{\perp}^z\equiv J_{\perp}=0.5$eV, 
while the intra-layer one for $3d_{z^2}$-orbital is $J_{\parallel}^x\equiv J_{\parallel}=0.25$eV.
The transmitted interaction strengths $J_{\perp}^x$ and $J_{\parallel}^{z\alpha}$ depend on the occupied densities of the relevant orbitals.

The $t$-$J$-like model is appropriate for strongly correlated systems where double occupancy is prohibited; SBMF is particularly well-suited for studying such models as its formalism naturally incorporates this constraint.
In the SBMF theory \cite{kotliar1988,lee2006htsc}, the electronic operators are expressed as a combination of spinon and holon part, i.e., $c_{i\sigma}^{\dagger}= f_{i\sigma}^{\dagger} h_i$.
The spin exchange interactions could be decomposed into spinon hopping and pairing channels, yielding the following mean-field expressions:
\begin{equation}
\begin{aligned}
\chi_{\parallel}^{s\alpha}
=&\frac{3}{8}J_{\parallel}^{s\alpha}
\langle f_{s\alpha\uparrow}^{\dagger}(j) f_{s\alpha\uparrow}(i)
+f_{s\alpha\downarrow}^{\dagger}(j) f_{s\alpha\downarrow}(i)\rangle,   \\
\Delta_{\parallel}^{s\alpha}=&\frac{3}{8}J_{\parallel}^{s\alpha} 
\langle f_{s\alpha\downarrow}(j) f_{s\alpha\uparrow}(i)
-f_{s\alpha\uparrow}(j) f_{s\alpha\downarrow}(i) \rangle,   \\
\chi_{\perp}^{s}
=&\frac{3}{8}J_{\perp}^{s}
\langle f_{s2\uparrow}^{\dagger}(i) f_{s1\uparrow}(i)
+f_{s2\downarrow}^{\dagger}(i) f_{s1\downarrow}(i)\rangle , \\
\Delta_{\perp}^{s}
=&\frac{3}{8}J_{\perp}^{s}
\langle f_{s2\downarrow}(i) f_{s1\uparrow}(i) 
-f_{s2\uparrow}(i) f_{s1\downarrow}(i) \rangle,
\end{aligned}
\label{eq:OPHoppPair}
\end{equation}
with $s=x,z$ for the $3d_{x^2-y^2},3d_{z^2}$ orbital, respectively.
The strengths of the ground state order parameters and holon densities are numerically solved,
as summarized in Tab.~\ref{tab:MForderParameter}.
In the relevant parameter regime $J_{\parallel}=0.25$eV and $J_{\perp}/J_{\parallel}=2$, the holon densities for the $3d_{x^2-y^2}$-orbital fall within the overdoped regime, 
$\delta_{x1}=0.402$ and $\delta_{x2}=0.523$, resulting in a reduction in its intra-layer pairing strengths.
In contrast, $3d_{z^2}$-orbital holon densities are $\delta_{z1}=0.171$ and $\delta_{z2}=0.332$, with the outer-layer one $\delta_{z1}$ closely approaching the optimal doping level of cuprates. 
This property suggests the potential for achieving SC in this orbital.

\begin{figure}[t!]
\centering
\includegraphics[width=0.48\textwidth]{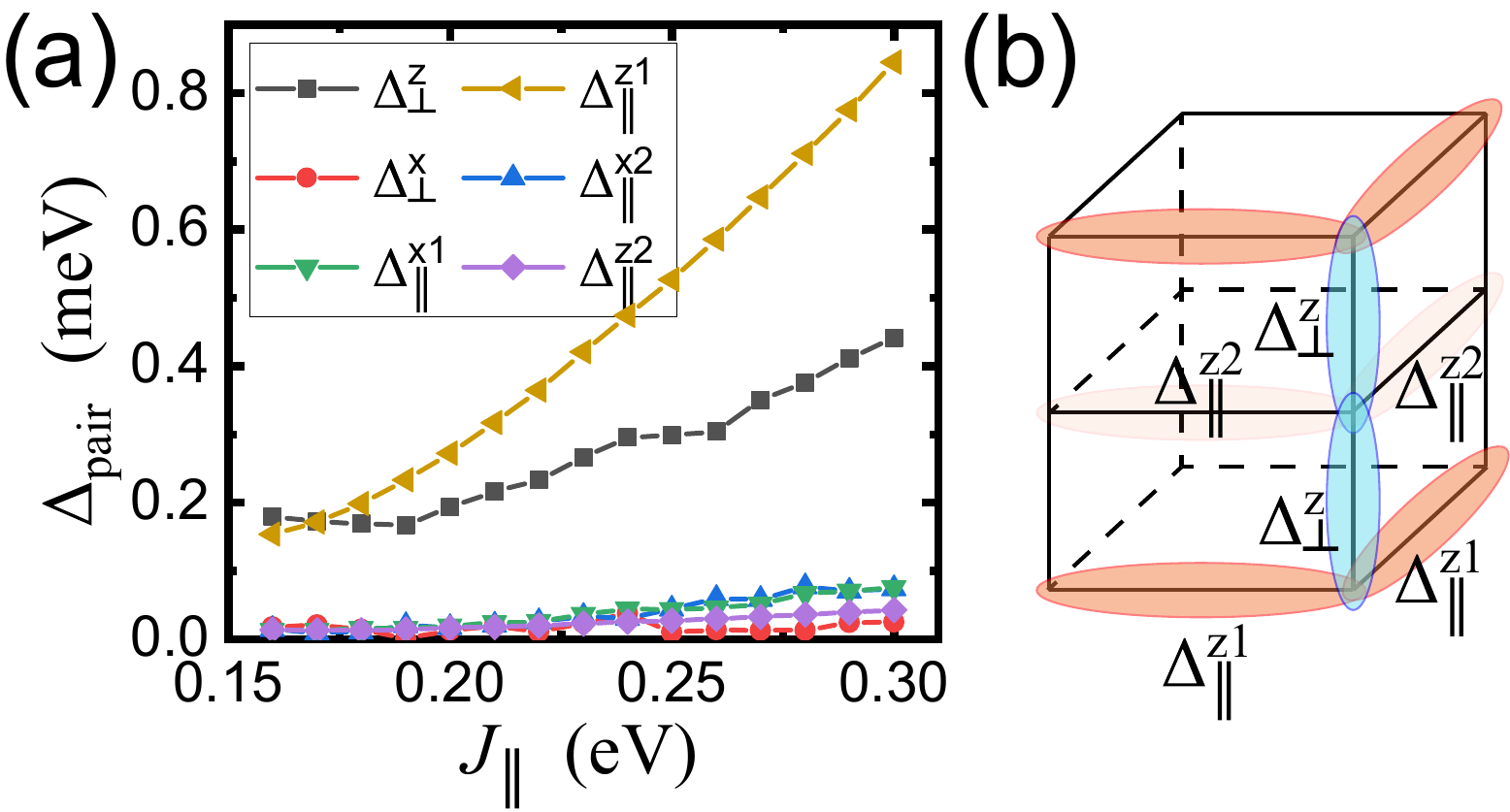}
\caption{(a) Spinon pairing strengths $|\Delta_{\mathrm{pair}}|$ versus exchange strengths $J_{\parallel}$ at fixed $J_{\perp}/J_{\parallel}=2$. 
(b) Schematic diagram for the dominant $3d_{z^2}$-orbital trilayer pairing structure. 
The intra-layer and inter-layer pairings exhibit opposite signs.}
\label{fig:SCPairing}
\end{figure}

{\bf Superconductivity:}
In the trilayer system, potential SC exhibits both intra-layer and inter-layer pairings for the two orbitals.
Numerical simulated results for the $J_{\parallel}$-dependence of all the relevant spinon pairing strengths are presented in Fig.~\ref{fig:SCPairing}(a).
Among the pairing channels, the inter-layer pairing $\Delta_{\perp}^z$ and outer-layer one $\Delta_{\parallel}^{z1}$ for the $3d_{z^2}$-orbital exhibit the largest amplitudes, being the dominance within the relevant parameter regime. 
Here, the intra-layer pairing exhibits the same phase along the $x$- and $y$- directions, leading to extended $s$-wave pairing.
Interestingly, the inter-layer and intra-layer pairings exhibit opposite signs, as illustrated in Fig.~\ref{fig:SCPairing}(b).
In a lateral view, the pairing character reflects a $d$-wave-like nature, signifying a change in the sign of the superconducting pairing from the parallel (intra-layer) to the perpendicular (inter-layer) direction.
The paring strengths in the $3d_{x^2-y^2}$ orbital are markedly suppressed owing to its overdoped nature.

\begin{figure}[t!]
\centering
\includegraphics[width=0.48\textwidth]{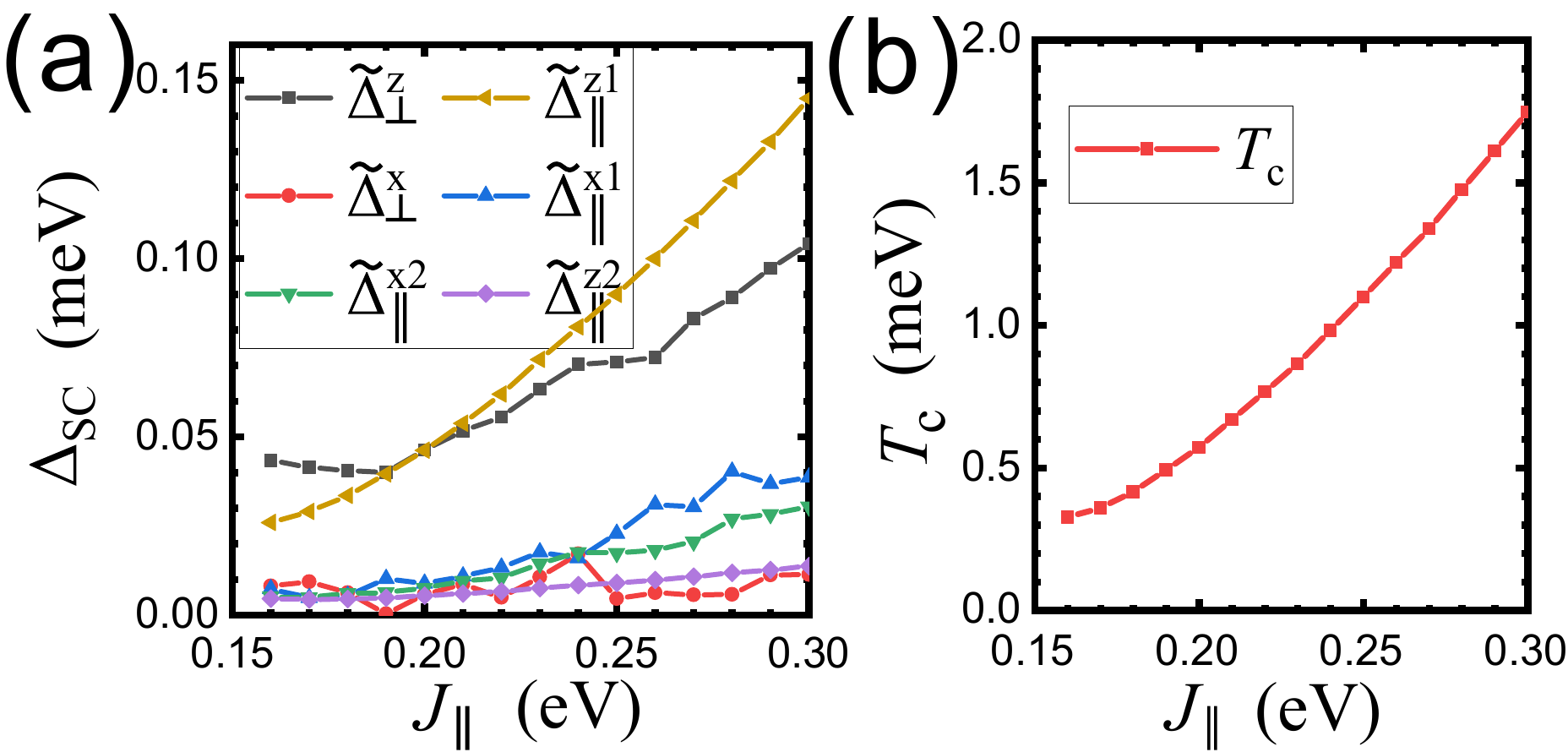}
\caption{(a) Superconducting pairing strengths $|\Delta_{\mathrm{SC}}|$ versus exchange strengths $J_{\parallel}$ with fixed $J_{\perp}/J_{\parallel}=2$. 
$3d_{z^2}$ orbital inter-layer pairing $\tilde{\Delta}_{\perp}^z$ and outer- extended $s$-wave pairing $\tilde{\Delta}_{\parallel}^{z1}$ dominate in the relevant parameter regime.
(b) Superconducting $T_c$ versus $J_{\parallel}$. 
}
\label{fig:SCTc}
\end{figure}

The superconducting state is realized when spinons are paired and holons are condensed \cite{kotliar1988,lee2006htsc}, 
characterized by the SC order parameter $\tilde{\Delta}_{\text{SC}}=\delta \Delta_{\text{pair}}$.
The numerical results for the $J_{\parallel}$ dependence of the several SC pairing strengths are summarized in Fig.~\ref{fig:SCTc}(a).
The superconducting $T_c$ is determined by the lower of the holon condensation temperature $T_{\text{BEC}}$ and the spinon pairing temperature $T_{\text{pair}}$. 
Holon condensation is effectively characterized by a generalized 2D XY-like model, 
where $T_{\text{BEC}}$ is replaced by the Kosterlitz–Thouless (KT) transition temperature \cite{kosterlitz1973kt}, 
proportional to the superfluid stiffness $\rho_{s\alpha}$ within orbital $s$ and layer $\alpha$.
The estimation of $\rho_{s\alpha}$ is given by:
\begin{equation}
\rho_{s\alpha}= 2 \delta_{s\alpha} t^{s\alpha}_{\parallel}  \chi^{s\alpha}_{\parallel} / (\frac{3}{8} J^{\alpha}_{s\parallel}).
\end{equation}
From the order parameters listed in Table.~\ref{tab:MForderParameter}, 
we roughly estimate $T_{\text{BEC}}^{s\alpha}=\frac{\pi}{2}\rho_{s\alpha}$ for the $s$ orbital in the $\alpha$ layer: $T_{\text{BEC}}^{x1}\approx 2.22 \times 10^{-1}$eV, $T_{\text{BEC}}^{x2}\approx 2.67 \times 10^{-1}$eV, $T_{\text{BEC}}^{z1}\approx 3.32 \times 10^{-3}$eV and $T_{\text{BEC}}^{z2}\approx 2.72 \times 10^{-3}$eV. 
Clearly, $3d_{x^2-y^2}$ orbital exhibits a much larger condensation temperature than $3d_{z^2}$ orbital, due to the larger holon densities.

Owing to the overdoped nature of the two orbitals, $T_c$ is governed by the spinon pairing temperature $T_{\text{pair}}$, which is lower than the condensation $T_{\text{BEC}}$. 
The dependence of superconducting $T_c$ on the spin exchange $J_{\parallel}$ with fixed $J_{\perp}/J_{\parallel}=2$ is depicted in Fig.~\ref{fig:SCTc}(b), exhibiting a qualitatively similar trend to the ground state pairing strength $\tilde{\Delta}_{\parallel}^{z1}$. 
In the presence of higher interaction strengths, $T_c$ naturally becomes larger with increasing $J_{\parallel}$.

The ground state order parameters reveal clear signatures of orbital-selective doping and correlation effects, highlighting the potential for SC to emerge predominantly in the $d_{z^2}$ orbital. 
Prior studies based on SBMF theory for the single-layer $t$-$J$ model have demonstrated that the spinon pairing temperature decreases with increasing doping, eventually becoming strongly suppressed in the heavily overdoped regime around $\delta\approx 0.3\sim 0.4$ \cite{kotliar1988}.
This theoretical trend aligns well with experimental observations in cuprates, where SC weakens beyond optimal doping \cite{kotliar1988,lee2006htsc,keimer2015highTc,proust2019highTc}.

Here, the numerical results for trilayer La$_4$Ni$_3$O$_{10}$ exhibit a consistent pattern. 
For the $d_{x^2-y^2}$ orbital, the hole concentration lies in the range of $\delta\approx 0.4\sim 0.5$, 
placing it deep in the overdoped regime. 
As a result, the pairing amplitude is significantly suppressed, and the corresponding spinon pairing temperature is very low. 
In contrast, the $d_{z^2}$ orbital remains less doped, with $\delta_{z1}=0.171$ for the outer layer and $\delta_{z2}=0.332$ for the inner layer. 
These lower doping levels support a stronger pairing amplitude and lead to a much higher pairing temperature relative to the $d_{x^2-y^2}$ orbital.

\begin{figure}[t!]
\centering
\includegraphics[width=0.48\textwidth]{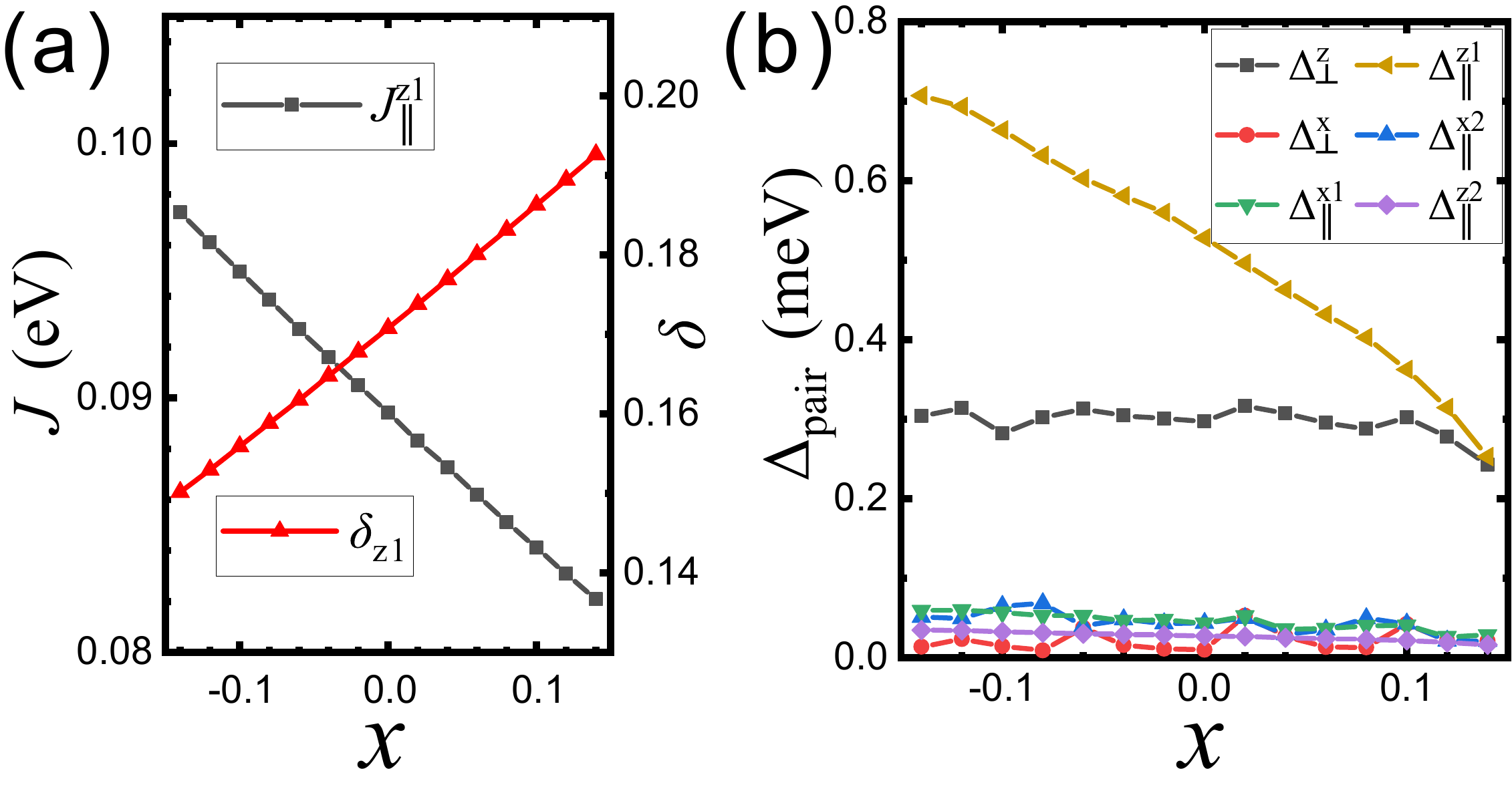}
\caption{(a) Outer-layer effective super-exchanges $J_{\parallel}^{z1}$ and holon densities $\delta_{z1}$ versus dopings $x$, with $x>0$ for hole dopings and $x<0$ for electron dopings.
(b) Spinon pairings versus dopings $x$. 
$3d_{z^2}$ orbital inter-layer pairing $\Delta_{\perp}^z$ nearly maintains the same amplitude under doping. 
}
\label{fig:DopingSCJ}
\end{figure}

{\bf Effect of doping:}
The doping dependence of the trilayer system is systematically investigated in Fig.~\ref{fig:DopingSCJ}, when the average number of electrons per unit cell is $4-x$ with doping fraction $x$.
Here, $x<0$ corresponds to electron doping, while $x>0$ corresponds to hole doping.
As the doping $x$ varies from $-0.1$ to $0.1$, the effective outer-layer $3d_{z^2}$-orbital holon density increases, as illustrated in Fig.~\ref{fig:DopingSCJ}(a). 
This is accompanied by a decrease in the effective coupling $J_{\parallel}^{z1}$, leading to a suppression in the out-layer spinon pairing $\Delta_{\parallel}^{z1}$\cite{kotliar1988}.
The simulated pairing strength $\Delta_{\parallel}^{z1}$ clearly exhibits an increase under electron doping and a reduction under hole doping, as depicted in Fig.~\ref{fig:DopingSCJ}(b).
The inter-layer pairings $\Delta_{\perp}^z$ of the $3d_{z^2}$-orbital maintain nearly the same amplitudes as $x$ varies.

\begin{figure}[t!]
\centering
\includegraphics[width=0.48\textwidth]{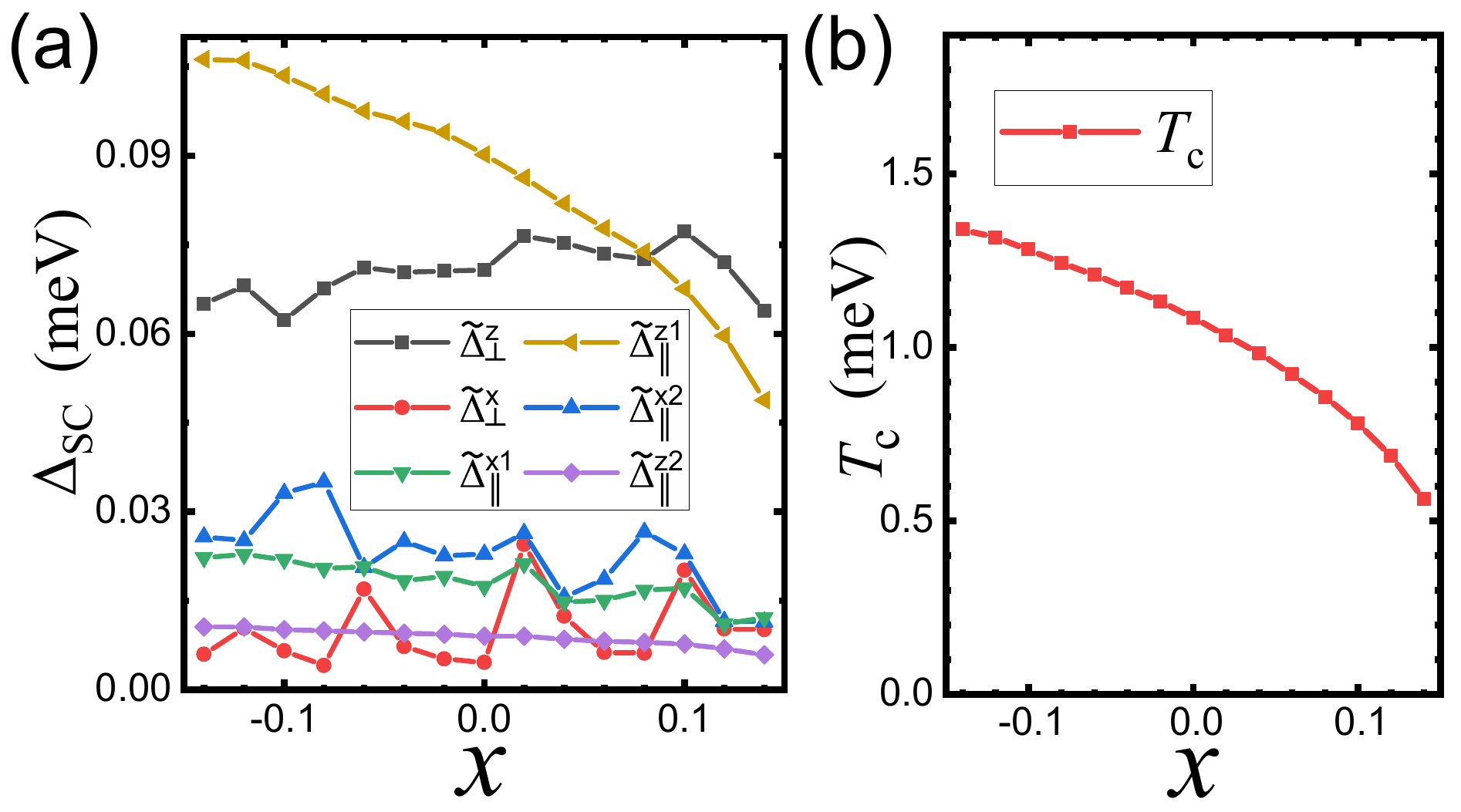}
\caption{
(a) Superconducting pairing order parameters versus dopings $x$. 
(b) Superconducting $T_c$ versus dopings $x$. 
$T_c$ increases under electron doping and reduces under hole doping.}
\label{fig:DopingSCTc}
\end{figure}

The evolution of the SC order parameters under doping is further verified in Fig.~\ref{fig:DopingSCTc}(a).
The inter-layer pairing and outer-layer pairing of the $3d_{z^2}$-orbital dominate.
Specifically, $\tilde{\Delta}_{\parallel}^{z1}$ is suppressed as $x$ increases.
The superconducting $T_c$ increase under the electron doping ($x<0$) and decrease under hole doping ($x>0$), 
as shown in Fig.~\ref{fig:DopingSCTc}(b), following a similar tendency as the $\tilde{\Delta}_{\parallel}^{z1}$.

{\bf Comparison between trilayer and bilayer system:}
The superconducting behaviors in La$_4$Ni$_3$O$_{10}$ and La$_3$Ni$_2$O$_{7}$ diverge due to differences in the $E_g$ orbital filling factor and multi-layer structure. 
In La$_3$Ni$_2$O$_{7}$, the bilayer configuration leads to a robust bonding band of $3d_{z^2}$ orbital, which is nearly fully occupied and localized \cite{lu2023bilayertJ,lu2023interplay}.
Consequently, the $3d_{x^2-y^2}$-orbital acts as the source of mobile carriers and exhibits a preference for strong inter-layer pairing.
Conversely, bands reconstruction of $E_g$-orbitals in trilayer La$_4$Ni$_3$O$_{10}$ is reduced.
As the $3d_{z^2}$ orbital is no longer close to half-filling, the effective coupling $J^{x}_{\perp}$ is weakened, as indicated in Eq.~(\ref{eq:EffTransJ}), leading to a significant reduction in inter-layer $3d_{x^2-y^2}$-orbital exchange.

\begin{figure}[t!]
\centering
\includegraphics[width=0.3\textwidth]{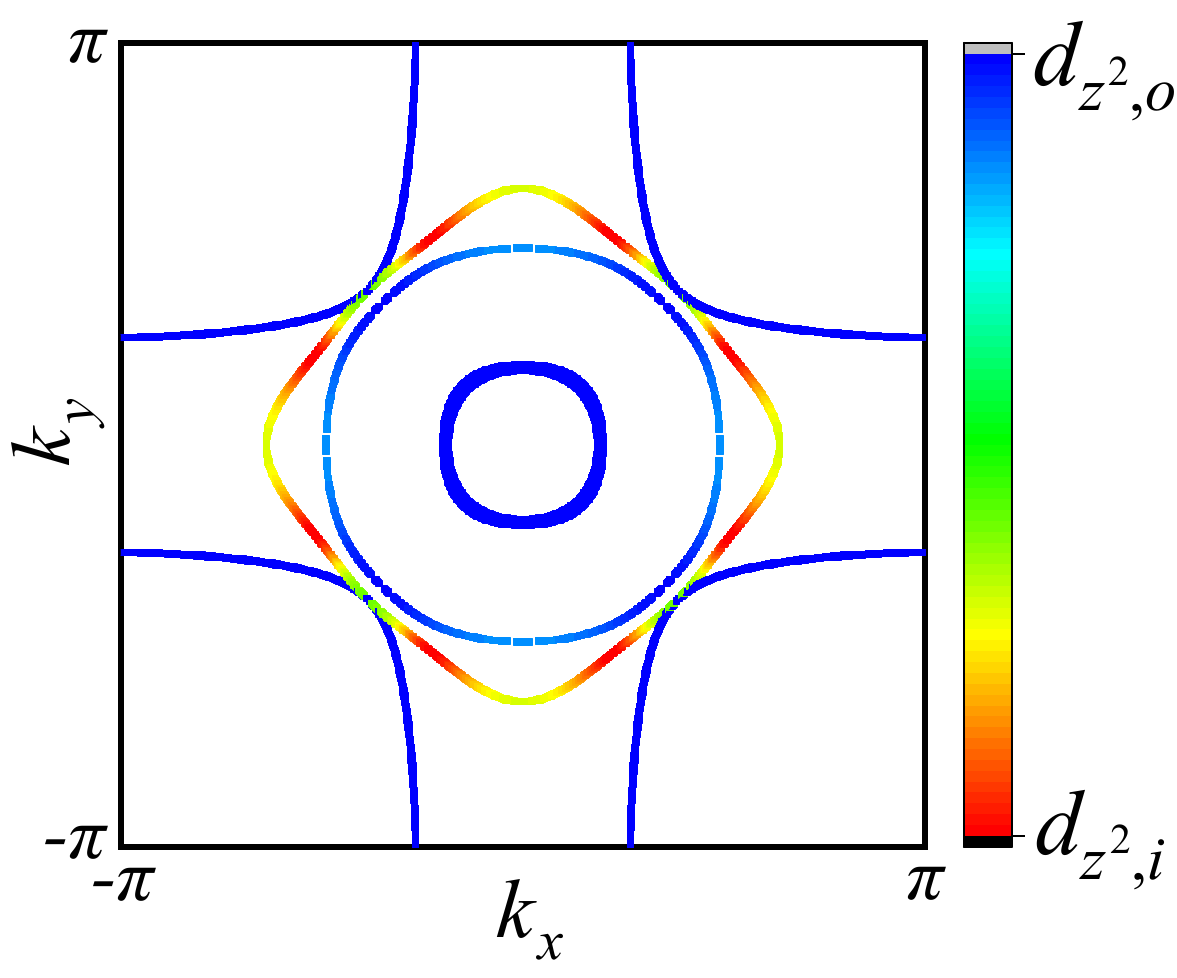}
\caption{The spinon Fermi-Surface. The right-side color panel shows the various components with the 
red and blue colors representing the inner-layer $3d_{z^2}$ and the outer-layer $3d_{z^2}$ orbitals, respectively.}
\label{fig:FS}
\end{figure}

The spinon Fermi-Surface (FS) in Fig.~\ref{fig:FS} further elucidates this distinction, where the color-coded compositions of the inner and outer layers are depicted. 
The outer-layer $3d_{z^2}$ orbital dominates in most of the regime, while the contribution from the inner-layer $3d_{z^2}$ orbital is minimal. 
The low density of the inner-layer $3d_{z^2}$ orbital near the FS in La$_{4}$Ni$_3$O$_{10}$ strongly hinders the inter-layer pairing of the $3d_{z^2}$ orbital compared to La$_3$Ni$_2$O$_{7}$. 
Moreover, the $3d_{z^2}$-orbital holon density in the outer layer is approximately $0.17$, close to the optimal hole-doping level. 
However, the intra-layer super-exchange interaction for $3d_{z^2}$ electrons is also weak, rendering intra-layer pairing fragile and susceptible to destruction by thermal fluctuations.
As a result, the superconducting behavior is predominantly characterized by intra-layer extended $s$-wave pairing in the outer layers combined with inter-layer pairing within $3d_{z^2}$ orbital.
Moreover, the pairing frustration induced by the trilayer structure and higher holon densities in $3d_{z^2}$ orbital in pressurized La$_4$Ni$_3$O$_{10}$ lead to a significantly lower $T_c$ compared to pressurized La$_3$Ni$_2$O$_{7}$.

The differences between our calculated spinon FS and ARPES measurements \cite{li2017fermiology,zhang2023trilayer,du2024correlated} likely stem from the simplifications in our tight-binding model, which deliberately excludes longer-range hopping and certain interactions to focus on dominant correlation effects. 
While including these additional terms could refine quantitative details like FS curvature and nesting features, 
we do not expect them to qualitatively alter the fundamental pairing symmetry or mechanism identified here.

{\bf Discussion:}
In summary, our investigation reveals that the $3d_{z^2}$ orbital primarily drives the superconducting behavior in the trilayer compound La$_4$Ni$_3$O$_{10}$, while $3d_{x^2-y^2}$ orbital plays an important role as a hidden bridge.
An intra-layer extended $s$-wave pairing within the outer layers, accompanied by an inter-layer pairing with opposite sign is stabilized in the numerical simulation, which exhibits a $d$-wave configuration in the side view. 
The numerical simulation further predicts that electron doping could enhance the pairing strength and critical temperature.

It is important to note, however, that the $T_c$ values calculated in our study are somewhat lower than those observed experimentally \cite{li2023trilayer,zhu2023trilayer,zhang2023trilayer}. 
This discrepancy can be primarily attributed to two main simplifications in our minimal model. 
First, by focusing on the most relevant degrees of freedom, 
our model neglects long-range hopping processes and additional interactions that could potentially enhance $T_c$. 
Secondly, the SBMF approximation neglects certain quantum and thermal fluctuations which can play a significant role in strongly correlated systems and influence the transition temperature. 
Despite these limitations, our approach captures the proposed pairing mechanism's qualitative features, including its orbital selectivity and doping dependence, thus providing valuable insights despite the difference in absolute $T_c$ values.

The discovery of superconductivity in pressurized La$_{n+1}$Ni$_n$O$_{3n+1}$ for $n=2$ and $n=3$ naturally raises the question of the optimal layer number, $n$, for SC within this nickelates series. 
Current experimental evidence indicates that $n=2$ already represents the optimal layered structure for these materials, 
in contrast to multilayer cuprates, where $n=3$ or $n=4$ is considered optimal for SC \cite{watanabe2017multicuprate}. 
This divergence may be attributed to differences in the underlying electronic structures: Nickelates La$_{n+1}$Ni$_n$O$_{3n+1}$ series have two $E_g$ orbitals near the Fermi level, while cuprates exhibit a single $3d_{x^2-y^2}$ orbital. 
Additionally, the $3d_{z^2}$ orbital in La$_{n+1}$Ni$_n$O$_{3n+1}$ facilitates strong interlayer hopping, resulting in robust interlayer superexchange, whereas cuprates only exhibit weak interlayer Josephson coupling \cite{ubbens1994,kuboki1995}. 
Furthermore, the multi-orbital nature and the role of Hundness \cite{georges2013hundness,tian2023correlation,qu2023roles,ouyang2023hund,kakoi2023pair} in generating strong correlations in nickelates may play a pivotal role in their superconductivity. 
A comprehensive understanding of the complex nature of superconductivity in this class of materials requires further theoretical and numerical investigations.

\section*{Acknowledgments} 
We are grateful to the stimulating discussions with Wei Li. 
C.W. is supported by the National Natural Science Foundation of China under the Grants No. 12234016 and No. 12174317.  
F.Y. is supported by the National Natural Science Foundation of China under the Grants No. 12074031. 
C.L. is supported by the National Natural Science Foundation of China under the Grants No. 12304180.
This work has been supported by the New Cornerstone Science Foundation.


\twocolumngrid
\bibliography{references}


\end{document}